\documentclass[final,3p,times,twocolumn]{elsarticle}

\usepackage{amssymb}
\usepackage{amsmath}
\usepackage{url}
\usepackage{graphicx}
\usepackage{subfig}
\usepackage{graphicx}
\usepackage{float}
\usepackage{placeins}

\usepackage{rotating}
\usepackage{orcidlink}
\usepackage{tikz}
\usepackage{hyperref}
\usepackage{enumitem}
\usepackage{stfloats}
\usepackage{amsthm}

\newtheorem{definition}{Definition}
\newtheorem{conjecture}{Conjectura}
\definecolor{orcidlogocol}{HTML}{A6CE39}

\newcommand{\orcidicon}{%
    \tikz[baseline=-0.5ex]\node[shape=circle,fill=orcidlogocol,inner sep=1pt] {\tiny\textsf{ID}};%
}

\newcommand{\orcid}[1]{\href{https://orcid.org/#1}{\orcidicon}}

\usepackage{booktabs}

\begin{document}

\begin{frontmatter}



\title{Topological Percolation in Urban Dengue Transmission: A Multi-Scale Analysis of Spatial Connectivity}


\author[1,2]{Marcílio Ferreira dos Santos}
\ead{marcilio.santos@ufpe.br}
\ead[url]{https://orcid.org/0000-0001-8724-0899}

\author[2,3]{Cleiton de Lima Ricardo}
\ead{cleiton.lricardo@ufpe.br}
\ead[url]{https://orcid.org/0000-0002-7114-1201}

\cortext[cor1]{Corresponding author.}


\affiliation[1]{
    organization={Núcleo de Formação de Docentes, Universidade Federal de Pernambuco (UFPE)}, 
    addressline={}, 
    city={Caruaru},
    postcode={},
    state={PE},
    country={Brazil}
}

\affiliation[2]{
    organization={Núcleo Interdisciplinar de Ciências Exatas e da Natureza (NICEN), Universidade Federal de Pernambuco (UFPE)}, 
    addressline={}, 
    city={Caruaru},
    postcode={},
    state={PE},
    country={Brazil}
}

\begin{abstract}
We investigate the spatial organization of dengue cases in the city of Recife, Brazil,
from 2015 to 2024, using tools from statistical physics and topological data analysis.
Reported cases are modeled as point clouds in a metric space, and their spatial
connectivity is studied through Vietoris--Rips filtrations and zero-dimensional
persistent homology, which captures the emergence and collapse of connected components
across spatial scales. By parametrizing the filtration using percentiles of the empirical
distance distribution, we identify critical percolation thresholds associated with
abrupt growth of the largest connected component. These thresholds define distinct
geometric regimes, ranging from fragmented spatial patterns to highly concentrated,
percolated structures. Remarkably, years with similar incidence levels exhibit
qualitatively different percolation behavior, demonstrating that case counts alone do
not determine the spatial organization of transmission. Our analysis further reveals
pronounced temporal heterogeneity in the percolation properties of dengue spread,
including a structural rupture in 2020 characterized by delayed or absent spatial
percolation. These findings highlight percolation-based topological observables as
physically interpretable and sensitive descriptors of urban epidemic structure,
offering a complementary perspective to traditional spatial and epidemiological analyses.
\end{abstract}

\begin{graphicalabstract}
\end{graphicalabstract}

\begin{highlights}[label=--]
\item We study the spatial organization of urban dengue cases using a percolation-based
topological framework grounded in zero-dimensional persistent homology.
\item Spatial connectivity is analyzed through Vietoris--Rips filtrations parametrized by
percentiles of the empirical distance distribution, enabling scale-free comparison across years.
\item Distinct percolation regimes are identified, ranging from fragmented spatial patterns
to highly concentrated, percolated structures.
\item Years with similar incidence levels exhibit markedly different percolation behavior,
demonstrating that case counts alone do not determine epidemic spatial structure.
\item A pronounced structural rupture is observed in 2020, characterized by delayed or
absent spatial percolation, consistent with altered urban mobility.
\end{highlights}

\begin{keyword}
Topological Data Analysis; Persistent Homology; Geometric Percolation;
Statistical Physics; Spatial Point Processes;
Urban Epidemics; Dengue Transmission
\end{keyword}

\end{frontmatter}




\section{Introduction}\label{sec:introduction}

Dengue remains one of the most impactful vector-borne diseases worldwide, with estimates indicating approximately 390 million infections annually and recurrent outbreaks across tropical and subtropical regions. Its primary vector, \textit{Aedes aegypti}, exhibits a strong dependence on urban spatial structure, operating predominantly over short distances and generating local transmission clusters whose fine-scale geometry is crucial for epidemiological surveillance \cite{barcellos2001, teixeira2009}. From a physical perspective, this dynamics suggests a spatial connectivity process governed by local interactions within a heterogeneous urban medium, in which macroscopic patterns emerge from restricted microscopic interactions.

Traditional epidemiological models, including compartmental frameworks and classical statistical approaches, describe aggregated trends but fail to recover the underlying geometry of transmission \cite{keeling2008}. Even more sophisticated spatial models often rely on coarse administrative units, which can smooth out or mask local heterogeneities that play a central role in vector--host interactions. In terms of statistical physics, this limitation corresponds to a loss of structural information associated with local connectivity and the formation of spatial clusters.

Topological Data Analysis (TDA) provides a complementary perspective that is particularly well suited to this challenge. Rather than imposing predefined spatial partitions, TDA extracts multiscale geometric properties directly from the observed point cloud, quantifying connectivity, fragmentation, and the emergence of topological structures associated with spatial heterogeneity. Persistent homology, the core tool of TDA, tracks the birth and death of these structures across filtration scales, producing invariants capable of summarizing complex patterns in a mathematically rigorous manner
\cite{edelsbrunner2000, zomorodian2005, chazal2016}.

From the viewpoint of complex systems physics, the evolution of connectivity along the filtration scale can be interpreted as a geometric percolation process in a continuous spatial system, in which local connected components progressively merge until the emergence of a dominant cluster \cite{stauffer1994, penrose2003}. Analogous processes have been widely studied in random geometric graphs, spatial networks, and urban systems, where the underlying geometry plays a central role in shaping global dynamics
\cite{barthelemy2011}.

Although TDA has been successfully applied in areas such as disordered materials, complex networks, granular systems, neuroscience, and climate sciences \cite{kramar2013}, its systematic use in spatial epidemiology remains incipient. In particular, few studies explicitly explore the relationship between persistent homology, spatial percolation, and structural connectivity regimes in high-resolution urban epidemiological data.

The availability of street-level georeferenced dengue data in Recife creates an unprecedented opportunity to construct ``epidemic shapes'' directly from the distribution of reported cases. Rather than explicitly modeling transmission chains, we examine the geometry of the epidemic field over time. Using Vietoris--Rips complexes constructed from Euclidean metrics in urban metric coordinates, we analyze zero-dimensional persistent homology, associated with the spatial connectivity of cases, across epidemiologically plausible scales.

We show that the evolution of spatial connectivity among dengue cases exhibits abrupt transitions consistent with distinct percolation regimes, enabling the classification of the analyzed years into patterns of concentrated, diffuse, and multifocal transmission. In particular, we identify critical scales associated with the topological collapse of spatial fragmentation, as well as atypical years whose topological signatures differ substantially from classical epidemic regimes. These results demonstrate that persistent homology captures structural aspects of dengue spatial dynamics that remain invisible to traditional spatial statistical approaches.

More broadly, this work illustrates how tools from statistical physics, percolation theory, and algebraic topology can be integrated to provide a geometric and multiscale characterization of disease spread in complex urban environments, positioning Topological Data Analysis as a promising framework for the
understanding of spatial epidemiological phenomena.

\section{Materials and Methods}
\label{sec:methods}

\subsection{Epidemiological data and geocoding}
\label{subsec:data_geocoding}

We used confirmed dengue case data reported in the municipality of Recife,
Brazil, between 2015 and 2024. Each record corresponds to an individual case
associated with a street address or postal code (CEP), from which geographic
coordinates were obtained using automated geocoding services. This procedure
assigns each address to an approximate location along the corresponding street
segment, typically near the centroid of the road axis \cite{dataset_recife_2025}.

It is well recognized that automated address geocoding does not provide the
exact physical location of a residence, but rather an estimate based on
interpolation along the street network. Classical studies in spatial
epidemiology have shown that this type of geocoding introduces unavoidable
positional error. In particular, comparisons between automatically geocoded
coordinates and true address locations obtained via visual inspection of aerial
imagery indicate that, in dense urban areas, the mean positional error is on the
order of 50--60~m, and that approximately 95\% of geocoded addresses lie within
150~m of their true locations \cite{cayo2003}.

Accordingly, the coordinates used in this study should be interpreted as
approximate spatial representations subject to positional uncertainty. While
this limitation may affect the precise location of individual cases, it does
not invalidate the analysis of collective and multiscale spatial patterns,
particularly those based on topological connectivity, which are known to be
stable under small perturbations of the underlying point cloud.

\subsection{Coordinate system and spatial metric}
\label{subsec:metric}

To ensure metric consistency in the spatial analyses, all geographic
coordinates were projected into the SIRGAS~2000 / UTM Zone~25S coordinate system,
allowing distances to be expressed directly in meters. The Euclidean distance
in the projected plane was adopted as the spatial metric, serving as an
effective approximation of geographic proximity between cases at the urban
scale.

More complex metrics, such as network-constrained distances along the street
network or anisotropic metrics induced by urban morphology, were not
considered in this version of the study. Assessing the impact of alternative
distance measures on topological connectivity constitutes a natural direction
for future work.

\subsection{Point cloud representation and Vietoris--Rips complexes}
\label{subsec:vr_complex}

For each year analyzed, dengue cases were represented as a spatial point cloud
\[
D = \{x_1, x_2, \dots, x_N\} \subset \mathbb{R}^2,
\]
where each point $x_i$ corresponds to the geocoded location of a reported case
in the urban plane.

From this point cloud, a geometric graph parametrized by a connectivity radius
$\varepsilon > 0$ was constructed, in which two points $x_i$ and $x_j$ are
connected by an edge whenever the Euclidean distance satisfies
$\|x_i - x_j\| \le \varepsilon$. The Vietoris--Rips complex $VR(\varepsilon)$ is
defined as the clique complex associated with this graph, such that any set of
$k+1$ mutually connected vertices defines a $k$-simplex.

The continuous variation of the parameter $\varepsilon$ induces an increasing
filtration of simplicial complexes,
\[
VR(\varepsilon_1) \subseteq VR(\varepsilon_2), \quad \varepsilon_1 < \varepsilon_2,
\]
which describes the multiscale evolution of spatial connectivity among cases.

\subsection{Zero-dimensional persistent homology}
\label{subsec:h0}

In this work, the analysis focuses on zero-dimensional persistent homology
($H_0$), which captures exclusively the connectivity structure of the system.
Each $H_0$ class corresponds to a connected component of the complex, whose birth
occurs when a point appears in isolation, and whose death corresponds to the
merging of that component with another as the parameter $\varepsilon$ increases.

The evolution of $H_0$ classes along the filtration provides a multiscale
description of the coalescence process among spatial clusters, allowing the
identification of fragmented regimes, intermediate phases, and the collapse of
connectivity into a dominant component, without imposing administrative
partitions or explicit parametric models.

\subsection{Topological observables and identification of transitions}
\label{subsec:observables}

For each value of $\varepsilon$, the following structural observables were
extracted:
\begin{itemize}
    \item the total number of connected components $N_c(\varepsilon)$;
    \item the size of the largest connected component $S_{\max}(\varepsilon)$;
    \item the distribution of sizes of intermediate components.
\end{itemize}

The critical connectivity scale $\varepsilon_{\mathrm{crit}}$ was identified
operationally as the value of $\varepsilon$ associated with the largest discrete
negative variation of $N_c(\varepsilon)$ along the filtration, corresponding to
the most abrupt structural reorganization of spatial connectivity. This behavior
is interpreted as an empirical transition analogous to a geometric percolation
process in finite spatial systems.

\subsection{Analyzed spatial scales}
\label{subsec:scales}

The Vietoris--Rips filtration was performed for values of $\varepsilon$ ranging
from 100~m to 500~m, with increments of 10~m. This interval was selected based on
epidemiological and urban considerations, reflecting the limited mobility of the
\textit{Aedes aegypti} vector and the typical organization of urban space into
blocks and micro-neighborhoods.

This multiscale sweep enables the identification of regions of topological
stability, as well as abrupt transitions in connectivity, without imposing a
priori a privileged spatial scale.

\subsection{Considerations on null models}
\label{subsec:null_models}

In the present version of the study, no systematic comparison with spatial null
models was performed, such as random point distributions constrained to the
urban support. Although such models are relevant for assessing the extent to
which the observed topological signatures exceed those expected from urban
geometry alone, this analysis is beyond the scope of the current version
(arXiv) and will be addressed in future work.

\section{Percolation structure and interpretation in finite systems}

Percolation theory provides a fundamental conceptual framework for understanding
emergent connectivity phenomena in spatial systems. Although originally developed
in the context of ideal disordered media and infinite networks, its core principles
have been extensively employed to interpret structural transitions in finite and
empirical systems, including geometric networks and real spatial processes.

In a geometric setting, percolation refers to the emergence of large-scale
connectivity as the interaction radius between spatially embedded points is
progressively increased.

\begin{definition}[Geometric percolation]
Given a finite set of points embedded in a metric space, geometric percolation
denotes the emergence of a macroscopic connected component as the interaction
radius $\varepsilon$ between points increases. This phenomenon is typically
associated with abrupt changes in observables related to the global connectivity
of the system.
\end{definition}

Classical results on random geometric graphs show that, even in purely random
spatial configurations, global connectivity does not arise gradually. Instead,
it emerges through a relatively abrupt transition associated with a critical
interaction scale. Below this scale, the system is dominated by small,
disconnected components; above it, a dominant component appears, concentrating a
significant fraction of the vertices.

In the theory of infinite systems, percolation is characterized by the existence
of a unique infinite connected component above a critical threshold, with
probability one. While this result does not apply literally to finite systems, it
motivates the use of the largest connected component as a dominant structural
observable in empirical analyses. In real spatial systems, the transition toward
a percolating regime is therefore expected to manifest as a rapid and concentrated
growth of the largest component, accompanied by a sharp reduction in the total
number of connected components. This form of structural dominance provides a clear
empirical signature of nontrivial spatial organization.

For finite systems, such as urban or epidemiological datasets, no phase
transition exists in the strict sense of the thermodynamic limit. Nevertheless,
statistical physics establishes that finite systems exhibit \emph{percolation
signatures}, expressed as rapid, though smooth, variations in connectivity
observables as a control parameter is varied. These finite-size signatures are
widely used to identify behaviors analogous to percolation in empirical data,
particularly when the system is strongly constrained by geometric or spatial
structure, as is the case in urban environments.

Within this framework, an operational definition of an empirical percolation
scale can be introduced.

\begin{definition}
\textbf{Definition (Empirical percolation scale).}
Let $N(\varepsilon)$ denote the number of connected components of the
Vietoris--Rips complex constructed from the point set for a given value of
$\varepsilon$. The empirical percolation scale $\varepsilon^*$ is defined as the
value of $\varepsilon$ associated with the largest negative variation of
$N(\varepsilon)$ along the filtration.
\end{definition}

This quantity provides an operational estimate of the scale at which the most
intense structural reorganization of the system occurs, acting as an empirical
order parameter for geometric percolation in finite datasets. Moreover, the
global shape of the fusion curve $N(\varepsilon)$ can be interpreted as an
aggregate indicator of the system’s connectivity dynamics, enabling systematic
comparisons across different spatial and temporal regimes.

\section{Topological Analysis of the Spatial Distribution of Dengue Cases (2015--2024)}

\subsection{Construction of topological indicators}

In this section, we analyze the spatial structure of dengue cases in the city of
Recife between 2015 and 2024 using tools from Topological Data Analysis (TDA),
with a focus on zero-dimensional persistent homology ($H_0$), which captures
spatial connectivity among georeferenced case records. This approach enables a
multiscale characterization of transmission patterns, going beyond purely
aggregated metrics, density maps, or local spatial autocorrelation analyses.

The analysis is based on the construction of Vietoris--Rips complexes from the
geographic coordinates of reported cases, using the Euclidean metric in a
projected coordinate system (UTM) to ensure metric consistency across the
spatial scales considered. Filtrations were performed for values of
$\varepsilon$ ranging from 100~m to 500~m, an interval compatible with
epidemiologically plausible scales of indirect spatial interaction, local
mobility, and urban neighborhood structure.

Table~\ref{tab:tda_summary} summarizes the main topological indicators extracted
for each year, including:

\begin{itemize}
    \item[(i)] the total number of points ($N$), corresponding to the number of unique reported cases;
    \item[(ii)] the critical scale $\varepsilon_{\text{crit}}$, defined as the value of $\varepsilon$ associated with the most abrupt growth of the largest connected component along the filtration;
    \item[(iii)] the size of the largest connected component at $\varepsilon = 300$~m;
    \item[(iv)] the number of connected components at this scale;
    \item[(v)] the maximum number of intermediate components observed throughout the filtration process;
    \item[(vi)] the collapse scale $\varepsilon_{\text{collapse}}$, defined as the smallest value of $\varepsilon$ for which the largest component contains at least 50\% of the points.
\end{itemize}

\begin{table*}[htbp]
\centering
\caption{Summary of topological indicators ($H_0$) describing the spatial
structure of dengue cases in Recife (2015--2024).}
\label{tab:tda_summary}
\begin{tabular}{c c c c c c c}
\hline
Year & $N$ & $\varepsilon_{\text{crit}}$ (m) &
$|\mathcal{C}_{\max}|_{\varepsilon=300}$ &
$N_{\text{comp}}(300)$ &
$N_{\text{mid}}^{\max}$ &
$\varepsilon_{\text{collapse}}$ (m) \\
\hline
2015 & 6165 & 220 & 5353 & 47  & 56 & 220 \\
2016 & 4969 & 240 & 3744 & 52  & 39 & 240 \\
2017 & 1225 & 410 &   73 & 224 & 11 & 410 \\
2018 & 1270 & 390 &   88 & 210 & 16 & 410 \\
2019 & 2647 & 450 & 1313 & 113 & 19 & 310 \\
2020 & 1416 & 500 &  259 & 188 & 12 & 500 \\
2021 & 3448 & 260 & 2498 & 65  & 39 & 260 \\
2022 & 1338 & 420 &   74 & 188 & 20 & 420 \\
2023 & 1670 & 450 &  200 & 174 & 19 & 450 \\
2024 & 3636 & 280 & 2544 & 55  & 40 & 280 \\
\hline
\end{tabular}
\end{table*}

In addition to the summarized indicators, Fig.~\ref{fig:h0_percolation} shows the
evolution of the number of connected components along the Vietoris--Rips
filtration for each analyzed year. The curves reveal distinct patterns of
spatial connectivity reorganization, allowing direct visualization of the
formation of dominant components and the emergence of contrasting topological
regimes over the study period. Structural similarities among subsets of these
curves suggest the existence of natural groupings of topological regimes across
different years, opening perspectives for future analyses based on functional
clustering or distance measures between connectivity curves.

\subsection{Spatial connectivity regimes and epidemiological interpretation}

The results reveal the existence of distinct spatial connectivity regimes over
the analyzed period, reflecting profound structural differences in the spatial
organization of dengue transmission. Such regimes are not directly inferable
from traditional incidence metrics, density maps, or isolated local spatial
autocorrelation analyses, as they emerge from the multiscale dynamics of
connectivity among reported cases.

Years associated with intense epidemic outbreaks, such as 2015, 2016, 2021, and
2024, exhibit reduced values of $\varepsilon_{\text{crit}}$ and
$\varepsilon_{\text{collapse}}$, on the order of 220--280~m, indicating early
collapse of spatial connectivity. In these years, the largest connected
component concentrates more than 70\% of reported cases at spatial scales below
300~m, characterizing a highly concentrated spatial organization of
transmission.

In contrast, years of lower epidemic intensity, such as 2017, 2018, and 2022,
remain fragmented across the entire range of analyzed scales. These years
display elevated values of $\varepsilon_{\text{crit}}$, close to or exceeding
390~m, reduced sizes of the largest component at $\varepsilon = 300$~m, and a
large number of connected components even at larger scales, indicating a regime
of diffuse spatial transmission marked by the persistence of multiple small and
weakly connected clusters.

Between these extremes, the years 2019 and 2023 define an intermediate regime,
characterized by the prolonged coexistence of multiple medium-sized components
along the filtration, followed by a relatively late collapse of spatial
connectivity. This pattern suggests a multifocal transmission process, in which
different urban regions contribute simultaneously to the spatial dynamics before
the consolidation of a dominant structure.

The year 2020 stands out as an atypical case. Despite presenting a substantial
number of reported cases, its topological signature differs markedly from that
of classical epidemic years. Both $\varepsilon_{\text{crit}}$ and
$\varepsilon_{\text{collapse}}$ occur only at large spatial scales, close to
500~m, and at $\varepsilon = 300$~m the largest component contains less than 20\%
of the reported cases. This result highlights that high incidence does not
necessarily imply high spatial concentration, an aspect not captured by analyses
based solely on case counts or density maps.

Taken together, these results demonstrate that the topological analysis of
spatial connectivity among dengue cases enables the identification of structural
transitions in transmission regimes over time, revealing patterns of
concentration, diffusion, and multifocality that remain invisible to traditional
spatial approaches. In this context, zero-dimensional persistent homology acts as
a multiscale geometric marker of dominant component formation, providing a solid
basis for deeper physical interpretations developed in the following section.

\section{Percolation analysis of spatial connectivity} \label{sec:percolation} 
\begin{figure*}[ht] 
\centering \includegraphics[width=0.85\linewidth]{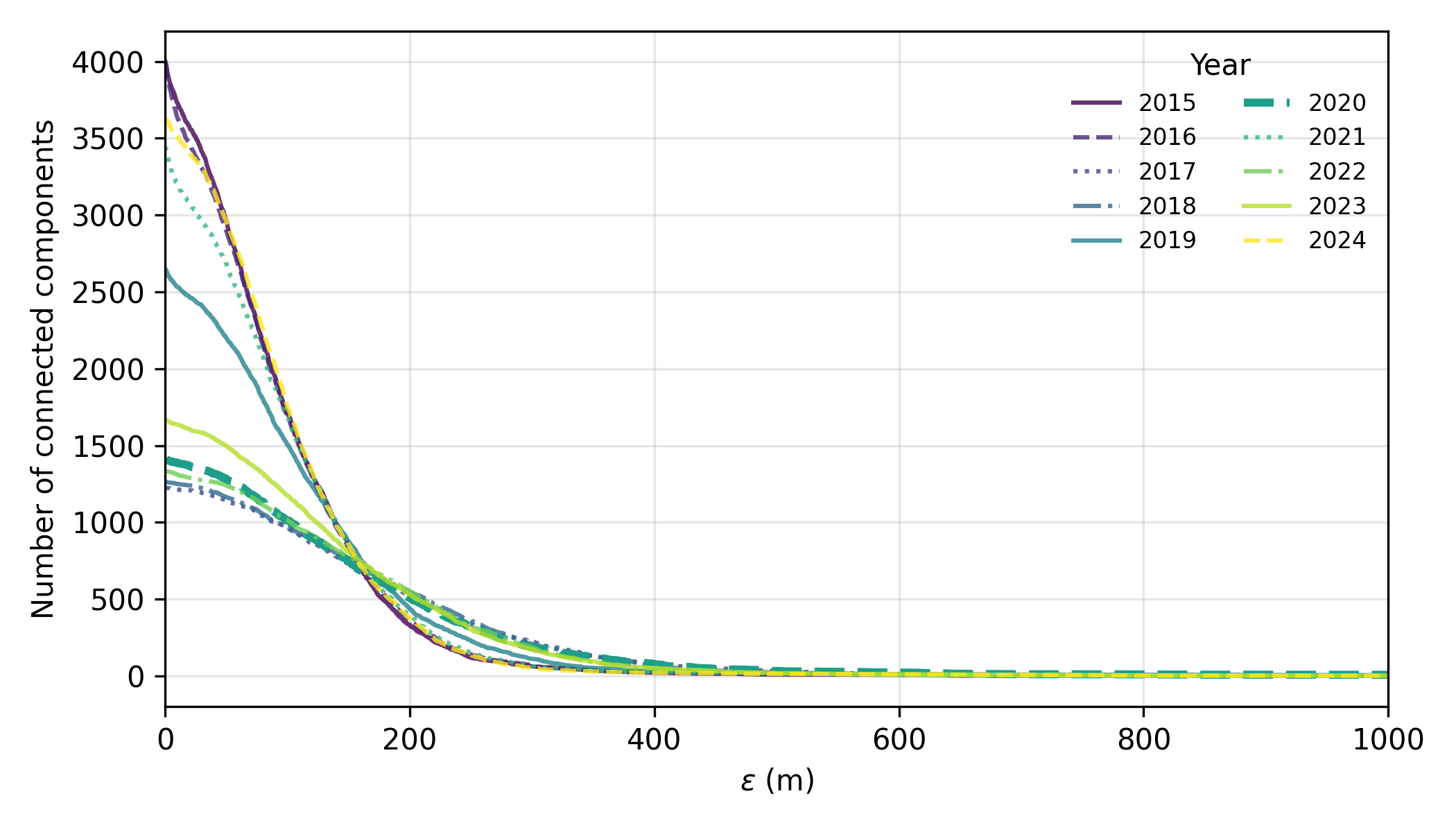} \caption{ Global fusion curves $N(\varepsilon)$ for the zero-dimensional persistent homology of dengue cases in Recife (2015--2024), restricted to $\varepsilon \leq 1$ km. Years with intense outbreaks exhibit rapid component coalescence at short spatial scales, while milder years show gradual and diffuse percolation behavior. } \label{fig:h0_percolation_all} 
\end{figure*}

To deepen the physical interpretation of the spatial patterns identified through zero-dimensional persistent homology ($H_0$), we analyze the spatial connectivity of dengue cases from the perspective of geometric percolation theory. This framework allows the organization of georeferenced cases to be interpreted as a continuous process of component coalescence driven by the spatial scale $\varepsilon$. Let $N(\varepsilon)$ denote the number of connected components of the Vietoris--Rips complex constructed from the metric coordinates of dengue cases. For $\varepsilon = 0$, each case forms an isolated component; as $\varepsilon$ increases, spatially proximate cases become connected, leading to the progressive merging of components and the eventual emergence of a giant connected component. The curve $N(\varepsilon)$ therefore provides a multiscale description of the spatial connectivity structure of the epidemic field. Figure~\ref{fig:h0_percolation_all} shows the global fusion curves for all years between 2015 and 2024, restricted to epidemiologically plausible spatial scales ($\varepsilon \leq 1$ km). Despite substantial interannual variation in the total number of cases, the curves exhibit well-defined percolative signatures that allow direct comparison between different spatial regimes. Years associated with intense dengue outbreaks display a rapid decay of $N(\varepsilon)$ at short spatial scales, indicating early formation of a giant component and characterizing a regime of fast percolation. Conversely, years with lower epidemic intensity show a gradual decrease of $N(\varepsilon)$ across a wide range of scales, reflecting persistent spatial fragmentation and the absence of a sharp percolation transition. 

\begin{figure*}[ht] 
\centering \includegraphics[width=0.48\linewidth]{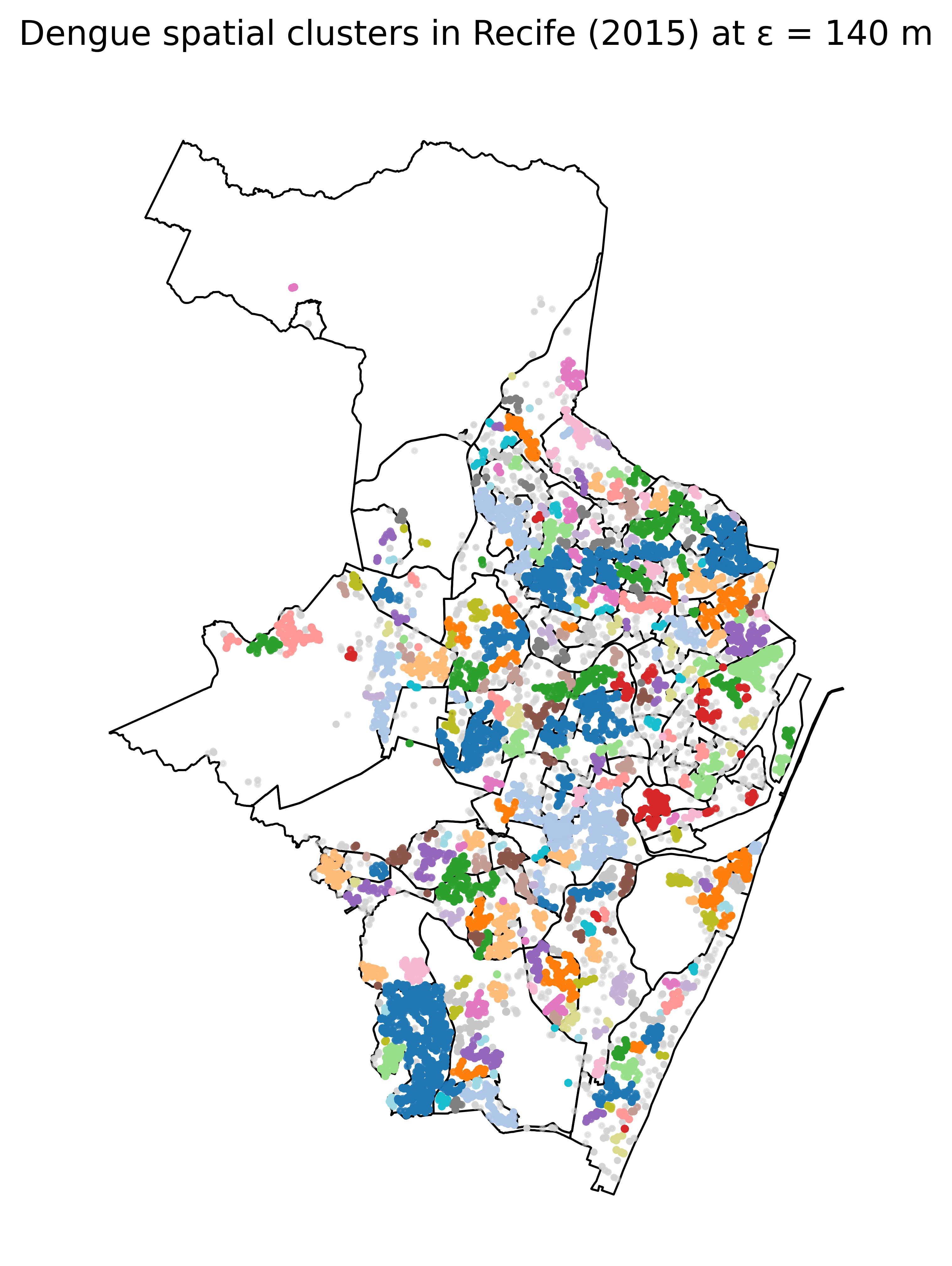} \includegraphics[width=0.48\linewidth]{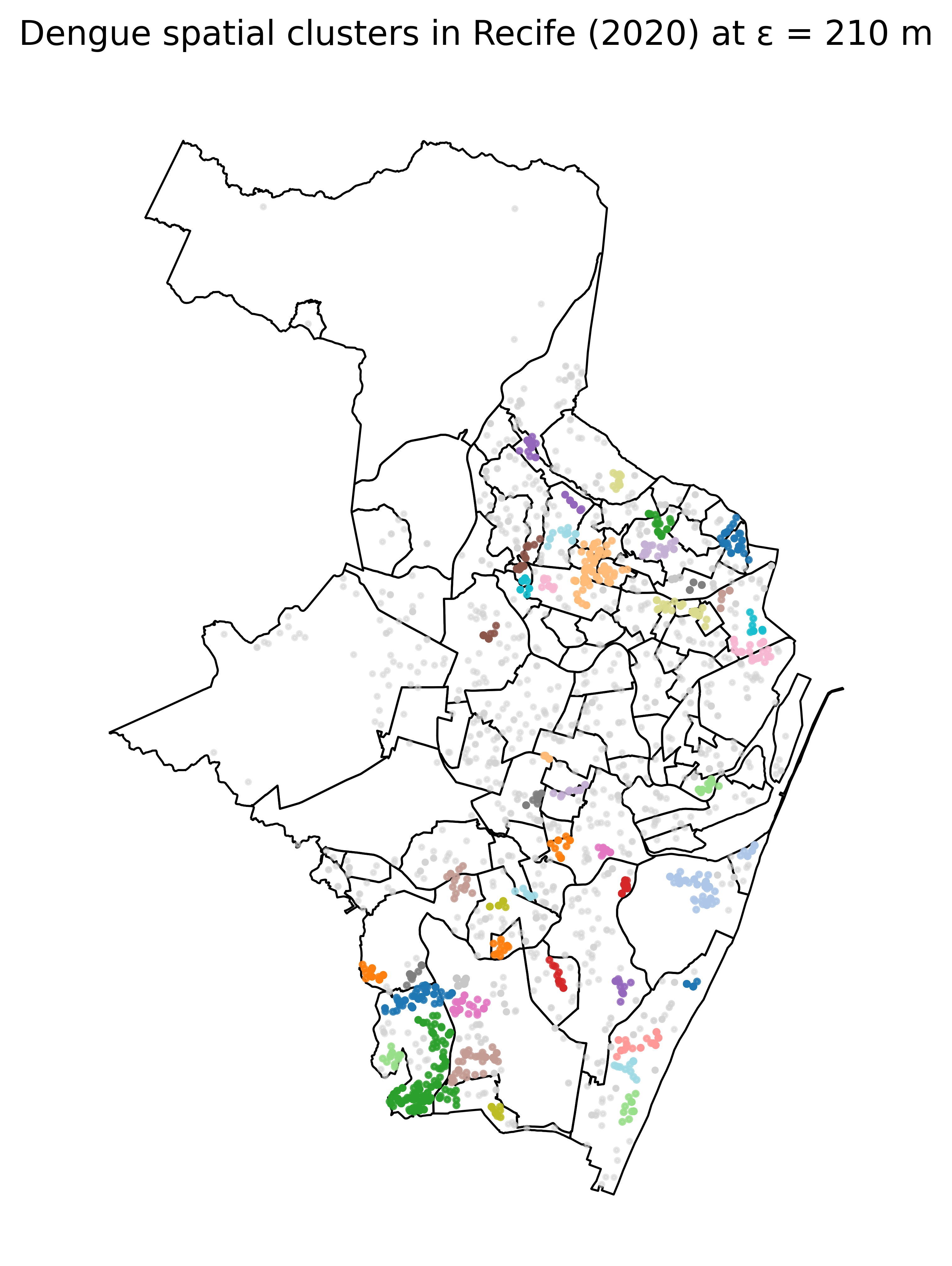} \caption{ Spatial distribution of connected components of dengue cases in Recife at the critical percolation scale $\varepsilon^*$. Left: 2015 ($\varepsilon^* = 140$ m), characterized by early emergence of large, compact clusters. Right: 2020 ($\varepsilon^* = 210$ m), exhibiting fragmented and spatially dispersed components. Colors indicate distinct connected components. } \label{fig:maps_eps} 
\end{figure*}

The percolative interpretation becomes particularly transparent when the global curves are contrasted with the explicit spatial distribution of connected components. Figure~\ref{fig:maps_eps} presents spatial maps of the connected clusters at the corresponding critical percolation scale $\varepsilon^*$ for two representative years. 

These maps confirm that rapid percolation is associated with spatially compact and strongly connected urban clusters, whereas delayed percolation corresponds to diffuse, multifocal transmission patterns.

\section{Percolation-based topological metrics and epidemic intensity}
\label{sec:percolation_metrics}

While persistent homology provides a detailed multiscale characterization of
spatial connectivity, its information can be synthesized into scalar observables
inspired by percolation theory. Such metrics condense the geometry of the fusion
process into interpretable quantities that capture the speed and abruptness of
spatial cluster coalescence, enabling direct comparison across years and
establishing a link between spatial organization and epidemic intensity.

\subsection{Critical percolation scale}

We define the \emph{critical percolation scale} $\varepsilon^*$ as the spatial
scale at which the rate of component merging reaches its maximum. Operationally,
$\varepsilon^*$ corresponds to the value of $\varepsilon$ that maximizes the
discrete derivative $-\Delta N / \Delta \varepsilon$ along the filtration.

From a physical perspective, $\varepsilon^*$ represents the characteristic
distance at which initially local clusters rapidly coalesce into a global
connected structure. Smaller values of $\varepsilon^*$ indicate faster spatial
percolation and a higher degree of spatial concentration of reported cases.

\subsection{Global percolation index}

To summarize the overall percolation behavior into a single scalar quantity, we
define the \emph{global percolation index} as
\begin{equation}
P = \int_0^{\varepsilon_{\max}} \frac{N(\varepsilon)}{N(0)} \, d\varepsilon,
\end{equation}
which is numerically approximated along the discrete filtration.

This quantity corresponds to the area under the normalized fusion curve and
provides an aggregate measure of the persistence of spatial fragmentation across
scales. Smaller values of $P$ indicate a rapid collapse of spatial connectivity,
whereas larger values reflect slow, diffuse, or weakly connected percolation
dynamics.

\subsection{Correlation with epidemic incidence}

To assess the epidemiological relevance of the percolation-based metrics, we
examined their association with annual dengue incidence. Spearman rank
correlation analysis reveals a strong and statistically significant negative
relationship between incidence and both observables:
\[
\rho(\varepsilon^*, I) \approx -0.99, \quad
\rho(P, I) \approx -0.98, \quad p < 10^{-3}.
\]

These results indicate that years with higher epidemic intensity tend to exhibit
faster spatial percolation and reduced persistence of spatial fragmentation.

\begin{figure*}[htbp]
    \centering
    \includegraphics[width=0.7\linewidth]{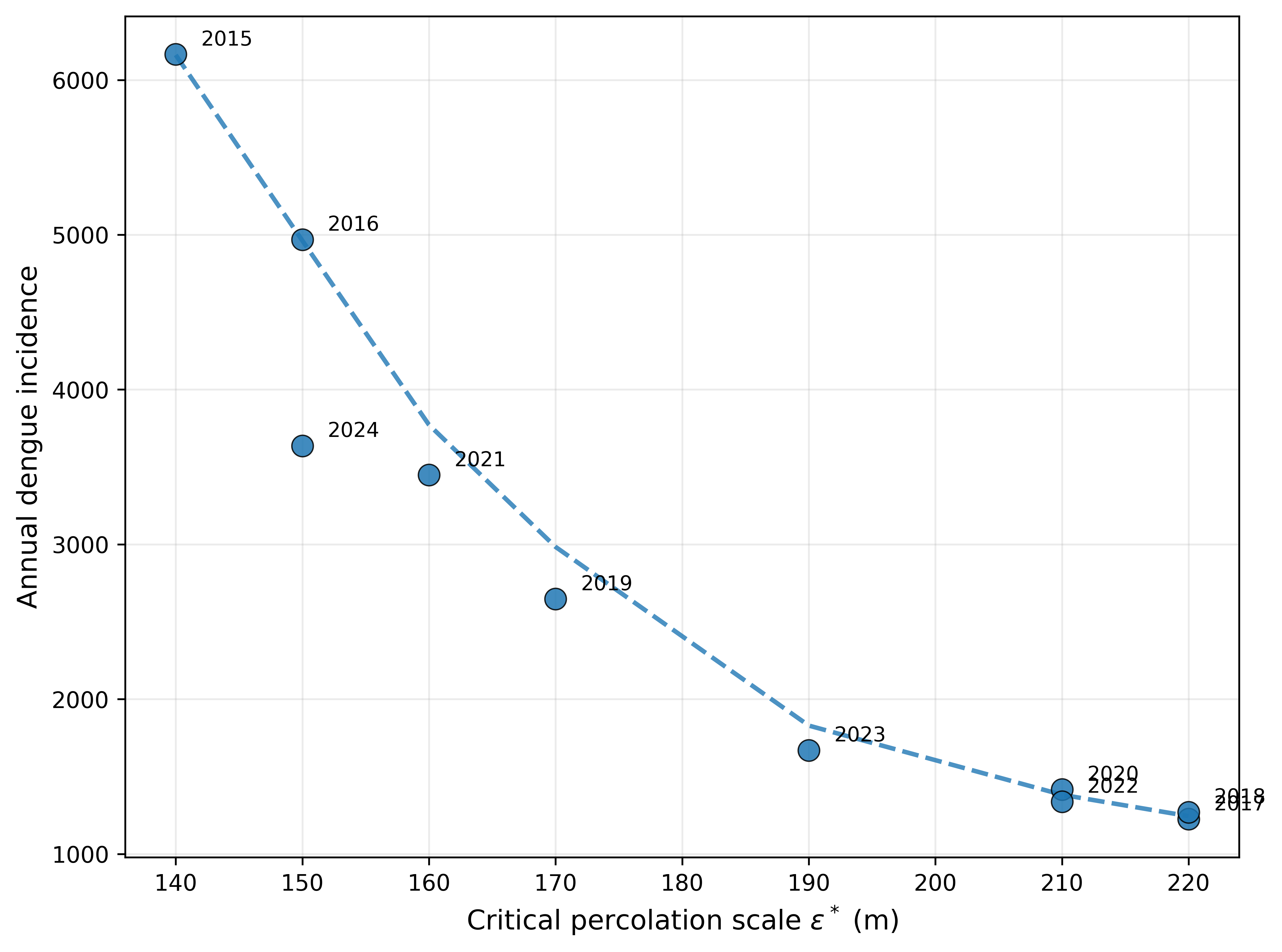}
    \caption{
    Relationship between the critical percolation scale $\varepsilon^*$ and annual
    dengue incidence in Recife (2015--2024). Lower values of $\varepsilon^*$
    indicate faster spatial percolation and are strongly associated with higher
    incidence levels. The dashed curve represents a LOWESS smoothing shown only as
    a guide to the eye.
    }
    \label{fig:scatter_eps_star}
\end{figure*}

\begin{figure*}[htbp]
    \centering
    \includegraphics[width=0.7\linewidth]{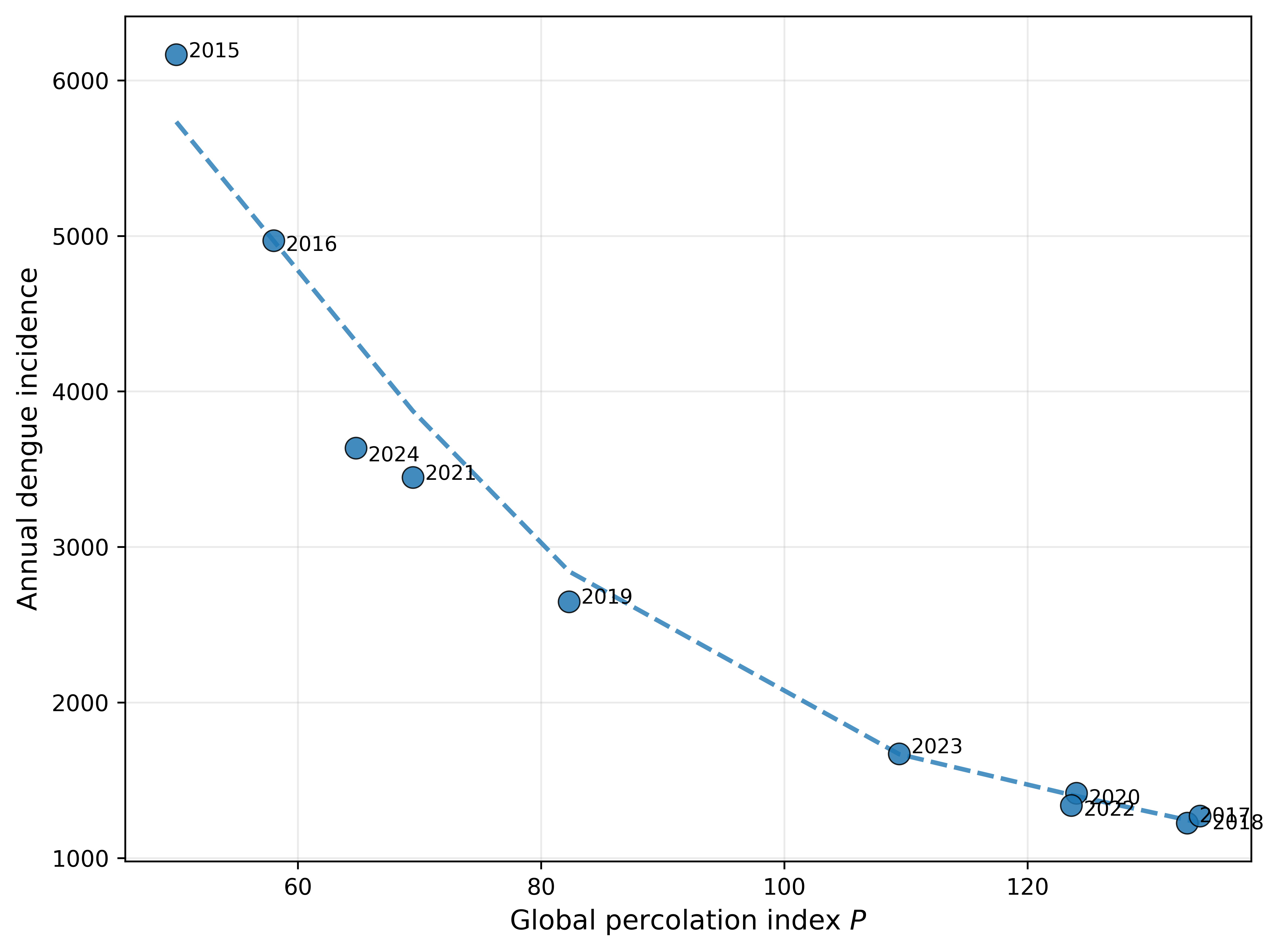}
    \caption{
    Relationship between the global percolation index $P$ and annual dengue
    incidence in Recife (2015--2024). Smaller values of $P$, corresponding to more
    rapid spatial percolation, are associated with increased epidemic intensity.
    The dashed curve represents a LOWESS smoothing shown only as a guide to the
    eye.
    }
    \label{fig:scatter_P}
\end{figure*}

\subsection{Physical interpretation and implications}

These results demonstrate that the spatial percolation properties of dengue case
distributions encode meaningful information about epidemic severity. The rapid
emergence of a giant connected component at short spatial scales reflects highly
concentrated urban transmission, whereas delayed percolation is associated with
diffuse or multifocal spatial dynamics.

Importantly, the metrics $\varepsilon^*$ and $P$ are derived exclusively from the
spatial geometry of the data, without imposing explicit transmission models or
assumptions regarding vector behavior. As such, they provide interpretable,
data-driven, and physically motivated descriptors of epidemic structure, with
potential applications in spatial surveillance and early warning systems.

\begin{conjecture}{Empirical rigidity of urban percolation}
In spatial processes constrained by an urban support, signatures of rapid
percolation characterized by small values of $\varepsilon^*$ and steep fusion
curves cannot be reproduced by null models that preserve only the underlying
spatial support. Such signatures therefore reflect non-random spatial
organization associated with structural mechanisms beyond simple urban geometry.
\end{conjecture}

This conjecture is supported by the empirical evidence presented in this work and
suggests that percolation-based topological observables may act as robust
detectors of emergent spatial structure in finite urban systems.

\section{Geometric Structure of Epidemic Connectivity}

The spatial dynamics of dengue transmission in urban environments constitute a finite complex system in which local interactions between hosts and vectors generate emergent collective patterns of connectivity. Although density maps and aggregated statistics provide information regarding epidemic intensity, such descriptors do not explicitly capture the underlying structural organization of spatial connectivity driving the transmission process.

In this work, we model the annual distribution of dengue cases as an empirical realization of a finite spatial point process and investigate its multiscale connectivity structure using tools from computational algebraic topology and geometric percolation theory.

\subsection{Geometric Representation of the Data}

Let
\[
\mathcal{D}_y = \{x_i\}_{i=1}^{N_y} \subset \mathbb{R}^2
\]
denote the set of spatial coordinates of the $N_y$ reported cases in year $y$, represented in the UTM coordinate system (meters), ensuring Euclidean metric consistency.

The spatial distance between two cases is defined as
\[
d_{ij} = \|x_i - x_j\|.
\]

This representation initially disregards administrative partitions, treating the data as a point cloud embedded in a continuous urban domain. Such a choice avoids spatial aggregation bias (the Modifiable Areal Unit Problem, MAUP) and enables a purely geometric analysis of connectivity.

\subsection{Vietoris–Rips Filtration}

To investigate multiscale connectivity, we employ the Vietoris–Rips filtration. For each scale parameter $\varepsilon \ge 0$, we define the simplicial complex

\[
\mathcal{VR}_\varepsilon(\mathcal{D}_y)
=
\left\{
\sigma \subset \mathcal{D}_y \;|\;
d_{ij} \le \varepsilon
\;\;\forall i,j \in \sigma
\right\}.
\]

In practice, we focus on zero-dimensional homology, which describes the number of connected components of the associated geometric graph. We define

\[
N_y(\varepsilon)
=
\text{number of connected components in }
\mathcal{VR}_\varepsilon(\mathcal{D}_y).
\]

For $\varepsilon = 0$, we have $N_y(0)=N_y$, since each point forms an isolated component. As $\varepsilon$ increases, components progressively merge until the eventual emergence of a dominant connected component.

The structural transition is characterized by the empirical critical scale

\[
\varepsilon^*_y
=
\arg\max_\varepsilon
\left(
-\frac{d N_y}{d \varepsilon}
\right),
\]

numerically estimated as the point of largest discrete negative variation of the curve $N_y(\varepsilon)$.

Physically, $\varepsilon^*_y$ represents the spatial scale at which the most abrupt component coalescence occurs, that is, the point of maximal structural reorganization of urban connectivity.

Large values of $\varepsilon^*$ indicate that a broader spatial radius is required to achieve global connectivity, suggesting a fragmented regime. Conversely, small values indicate early formation of a dominant component, suggesting spatial concentration of cases.

Beyond the critical scale, we define the global percolation index

\[
P_y
=
\int_{0}^{\varepsilon_{\max}}
\frac{N_y(\varepsilon)}{N_y(0)} \, d\varepsilon.
\]

This quantity measures the overall persistence of fragmentation across spatial scales.

High values of $P_y$ indicate that the network remains fragmented over a wide range of scales, whereas low values indicate rapid consolidation of connectivity.

While $\varepsilon^*$ captures the transition point, $P_y$ summarizes the global intensity of fragmentation.

\subsection{Interpretation as a Finite Percolation System}

It is important to emphasize that the system under investigation is finite and non-stationary. Unlike classical percolation on infinite lattices, we analyze an empirically bounded spatial realization.

Nevertheless, the behavior of the curve $N_y(\varepsilon)$ exhibits qualitative signatures analogous to percolation transitions in random geometric graphs: a fragmented phase, a rapid coalescence regime, and the emergence of a dominant component.

However, as demonstrated in subsequent sections, the observed organization cannot be explained solely by generic geometric properties, requiring validation against structured null models.

The choice of zero-dimensional homology is deliberate. In urban epidemiological systems, the spatial connectivity of infection foci constitutes a macroscopic variable relevant to collective propagation.

Rather than explicitly modeling vector dynamics, this approach extracts structural invariants directly from empirical geometry, enabling:

\begin{itemize}[label=-]
\item Interannual comparison independent of absolute incidence;
\item Identification of distinct structural regimes;
\item Multiscale analysis without arbitrary spatial partitioning;
\item Integration with complex systems theory and geometric percolation.
\end{itemize}

This formulation establishes the fundamental geometric framework upon which we introduce, in subsequent sections, temporal constraints and structured null model validation.

\section{Spatiotemporal Structure: Sliding Windows and Causal Filtration}

The purely spatial representation introduced in the previous section captures the aggregated annual geometry of the epidemic. However, dengue transmission is intrinsically spatiotemporal: epidemiologically plausible connections depend not only on spatial proximity but also on temporal compatibility between cases.

In this section, we introduce two distinct strategies for incorporating the temporal dimension and analyze their structural implications.

Let
\[
\mathcal{D}_y = \{(x_i, t_i)\}_{i=1}^{N_y},
\]
where $x_i \in \mathbb{R}^2$ denotes the spatial position (UTM coordinates) and
$t_i \in [0,T_y]$ the time in days within year $y$.

We define the temporal separation between two cases as
\[
\Delta t_{ij} = |t_i - t_j|.
\]

We fix a temporal horizon $\Delta T > 0$ (in days), interpreted as an epidemiologically plausible interaction window mediated by the vector.

We investigate two distinct constructions.

\subsection{Model A: Sliding Temporal Windows}

In the window-based model, the annual dataset is partitioned into independent temporal subsets.

For a fixed duration $\Delta T$, we define windows

\[
W_k
=
\left\{
(x_i,t_i) \in \mathcal{D}_y
\;|\;
t_i \in [\tau_k, \tau_k + \Delta T)
\right\},
\]

where $\tau_k$ traverses the year with fixed step size.

Within each window $W_k$, we apply a purely spatial Vietoris–Rips filtration:

\[
\mathcal{VR}_\varepsilon(W_k)
=
\left\{
\sigma \subset W_k
\;|\;
\|x_i - x_j\| \le \varepsilon
\;\forall i,j \in \sigma
\right\}.
\]

Let $N_k(\varepsilon)$ denote the number of connected components in window $k$.

The annual averaged curve is defined as

\[
N_{\text{window}}(\varepsilon)
=
\frac{1}{K}
\sum_{k=1}^{K}
N_k(\varepsilon).
\]

\paragraph{Structural Interpretation}

This model preserves local spatial structure within each temporal interval but eliminates structural continuity across distinct windows. Components that could potentially connect through temporal chaining are artificially disconnected.

From a complex systems perspective, the window-based model performs a temporal coarse-graining with periodic structural reinitialization.

\subsection{Model B: Weak Causal Filtration}

In the weak causal model, the full annual structure is preserved, but connections are allowed only when spatially and temporally compatible.

We define the masked metric

\[
\tilde{d}_{ij}^{(\Delta T)}
=
\begin{cases}
\|x_i - x_j\|,
& \text{if } \Delta t_{ij} \le \Delta T, \\
+\infty,
& \text{otherwise}.
\end{cases}
\]

The causal filtration is then defined as

\[
\mathcal{VR}_\varepsilon^{\text{causal}}
=
\left\{
\sigma \subset \mathcal{D}_y
\;|\;
\tilde{d}_{ij}^{(\Delta T)} \le \varepsilon
\;\forall i,j \in \sigma
\right\}.
\]

We denote by $N_{\text{causal}}(\varepsilon)$ the associated number of connected components.

\paragraph{Structural Interpretation}

Unlike the window-based model, no artificial partitioning of the year is introduced. Connections may occur between any two cases provided they satisfy the temporal constraint $\Delta T$.

This allows the formation of spatiotemporal chains that propagate across the entire year through successive local linkages.

The term “causal” is used in a structurally weak sense: we do not explicitly model vector dynamics, but we impose epidemiologically plausible temporal coherence.

\subsection{Fundamental Structural Difference}

The two constructions differ fundamentally:

\begin{itemize}[label=-]
\item The window-based model fragments annual structural continuity.
\item The causal model preserves global connectivity subject to local temporal constraints.
\end{itemize}

In terms of dynamic geometric graphs:

\begin{itemize}[label=-]
\item The window-based model corresponds to a sequence of independent graphs.
\item The causal model corresponds to a single temporally constrained spatiotemporal graph.
\end{itemize}

This distinction is central. If the observed organization arises solely from local spatial density, both models should produce similar structural signatures.

If genuine spatiotemporal chaining exists, the causal model is expected to exhibit:

\begin{itemize}[label=-]
\item a smaller critical scale $\varepsilon^*$;
\item a lower global percolation index $P$;
\item stronger contrast relative to null models that destroy spatiotemporal correlation.
\end{itemize}

The next section formalizes the null models employed to evaluate this hypothesis.

\section{Structural Comparison Between Filtering Pipelines and Validation via Null Models}
\label{sec:comparison_nulls}

In this section, we investigate the spatiotemporal organization of epidemic connectivity under two distinct filtering schemes:

\begin{enumerate}
\item Sliding temporal window filtration;
\item Weak causal filtration with global temporal constraint.
\end{enumerate}

The analysis is conducted under a fixed temporal horizon $\Delta T = 14$ days, chosen to reflect an epidemiologically plausible scale of infectious chaining.

Let the annual dataset be defined as

\[
\mathcal{D}_y = \{(x_i,t_i)\}_{i=1}^{N_y},
\]

where $x_i \in \mathbb{R}^2$ denotes the spatial position (UTM coordinates) and $t_i$ the time in days within year $y$.

Spatial distance is defined by

\[
d_{ij} = \|x_i - x_j\|.
\]

The structural metrics considered are:

\begin{itemize}[label=-]
\item the critical scale $\varepsilon^*$ (largest discrete drop of $N(\varepsilon)$);
\item the global percolation index
\[
P = \int \frac{N(\varepsilon)}{N(0)}\, d\varepsilon.
\]
\end{itemize}

\subsection{Filtering Pipelines}

\subsubsection*{Sliding Window Filtration}

The annual dataset is partitioned into independent windows of width $\Delta T$:

\[
W_k = \{ (x_i,t_i) \in \mathcal{D}_y \mid t_i \in [\tau_k,\tau_k+\Delta T) \}.
\]

Within each window, a purely spatial Vietoris–Rips filtration is applied, producing $N_k(\varepsilon)$.

The aggregated curve is defined as

\[
N_{\text{window}}(\varepsilon)
=
\frac{1}{K}\sum_{k=1}^{K} N_k(\varepsilon).
\]

This pipeline preserves local spatial structure but eliminates inter-window structural continuity.

\subsubsection*{Weak Causal Filtration}

Define temporal separation as

\[
\Delta t_{ij} = |t_i - t_j|.
\]

The masked metric is defined by

\[
\tilde{d}_{ij}^{(\Delta T)} =
\begin{cases}
d_{ij}, & \text{if } \Delta t_{ij} \leq \Delta T, \\
+\infty, & \text{otherwise}.
\end{cases}
\]

The causal filtration is constructed using $\tilde{d}_{ij}^{(\Delta T)}$, retaining the complete annual dataset.

This scheme preserves global structural continuity across the year, removing only temporally incompatible connections.

\subsection{Null Models}

Two complementary null models were employed.

\subsubsection*{Null Model A — Temporal Permutation}

Time labels are randomly permuted:

\[
t_i^{(\pi)} = t_{\pi(i)},
\]

preserving the marginal spatial and temporal distributions while destroying spatiotemporal correlation.

A total of 500 permutations were performed per year.

\subsubsection*{Null Model B — Inhomogeneous Spatial Resampling}

The estimated spatial intensity $\lambda(x)$ is preserved, and a new independent sample of $N_y$ points is generated according to this intensity.

This model maintains global urban heterogeneity but removes specific relational chaining.

\subsection{Comparative Results for $\Delta T = 14$ Days}

\subsubsection*{Causal Pipeline Under Two Null Models}

\begin{table}[h!]
\centering
\caption{Causal pipeline ($\Delta T = 14$ days) compared to two null models.}
\label{tab:causal_dual_null}
\begin{tabular}{cccccc}
\hline
Year & $\varepsilon^*$ & $P$ & $p_A(\varepsilon^*)$ & $p_B(\varepsilon^*)$ & $p_B(P)$ \\
\hline
2015 & 103.76 & 153.80 & 0.446 & 0.010 & 0.023 \\
2016 & 79.32  & 195.31 & 0.066 & 0.000 & 0.010 \\
2017 & 582.71 & 578.99 & 0.888 & 0.570 & 0.000 \\
2018 & 499.62 & 572.50 & 0.794 & 0.347 & 0.000 \\
2019 & 113.53 & 322.78 & 0.004 & 0.000 & 0.000 \\
2020 & 250.38 & 491.72 & 0.162 & 0.060 & 0.000 \\
2021 & 191.73 & 259.02 & 0.208 & 0.423 & 0.000 \\
2022 & 455.64 & 540.61 & 0.890 & 0.233 & 0.000 \\
2023 & 445.86 & 478.09 & 0.642 & 0.293 & 0.000 \\
2024 & 211.28 & 243.36 & 0.924 & 0.453 & 0.000 \\
\hline
\end{tabular}
\end{table}

We observe that:

\begin{itemize}[label=-]
\item Under temporal permutation (Null A), statistical significance varies across years;
\item Under the inhomogeneous spatial model (Null B), the global index $P$ remains systematically significant;
\item The critical scale $\varepsilon^*$ is partially explained by spatial heterogeneity, but not fully.
\end{itemize}

These results indicate that the observed structure cannot be attributed solely to urban density patterns.

\subsubsection*{Window-Based Pipeline Under Temporal Permutation}

\begin{table}[h!]
\centering
\caption{Sliding window pipeline compared to the temporal permutation null model ($\Delta T = 14$ days).}
\label{tab:window_temporal_null}
\begin{tabular}{cccccc}
\hline
Year & $\varepsilon^*$ & $P$ & $p(\varepsilon^*)$ & $p(P)$ \\
\hline
2015 & 100.84 & 276.05 & 0.592 & 1.000 \\
2016 & 100.84 & 329.57 & 0.528 & 1.000 \\
2017 & 605.04 & 942.03 & 0.600 & 0.004 \\
2018 & 789.92 & 971.66 & 0.858 & 0.824 \\
2019 & 100.84 & 552.01 & 0.000 & 1.000 \\
2020 & 605.04 & 763.14 & 0.982 & 1.000 \\
2021 & 100.84 & 446.95 & 0.252 & 1.000 \\
2022 & 689.08 & 904.94 & 0.780 & 0.048 \\
2023 & 487.39 & 827.49 & 0.412 & 0.078 \\
2024 & 201.68 & 447.78 & 0.226 & 0.992 \\
\hline
\end{tabular}
\end{table}

Comparatively, the window-based pipeline exhibits reduced discriminative power under temporal randomization.

Annual fragmentation diminishes statistical sensitivity to global spatiotemporal structural coherence.

The results suggest that:

\begin{itemize}[label=-]
\item A portion of the observed organization arises from underlying urban spatial heterogeneity;
\item However, the global percolation index remains significantly distinct even under spatially inhomogeneous control;
\item The causal filtration preserves multiscale structural coherence that is not captured by the window-based model.
\end{itemize}

These findings indicate that the identified organization constitutes a robust structural property of the urban transmission system, rather than a mere artifact of density effects or temporal randomness.

\section{Morphology of the Percolation Transition: Structural Shape Metrics}

Classical percolation metrics — the critical scale $\varepsilon^*$ and the integrated index $P$ — capture, respectively, the transition point and the aggregated intensity of connectivity. However, these quantities do not distinguish different \emph{transition morphologies}, that is, distinct functional shapes of the merging curve $N(\varepsilon)$.

Empirical observations indicate that years with markedly different epidemic intensities may exhibit similar values of $\varepsilon^*$, yet display globally distinct concavity patterns in their merging curves. In particular, we identify two qualitative regimes: (i) abrupt transitions, characterized by strong global upward concavity and associated with rapid structural collapse; and (ii) smooth, diffuse transitions, characterized by reduced or negative global concavity and associated with progressive connectivity.

To quantify these structural differences, we introduce a set of metrics derived from the discrete geometry of the merging curve.

Let $N(\varepsilon_k)$ denote the curve evaluated on an increasing grid $\{\varepsilon_k\}_{k=1}^K$. We define the discrete second difference:

\[
\kappa_k
=
N(\varepsilon_{k+1})
- 2N(\varepsilon_k)
+ N(\varepsilon_{k-1}),
\quad k=2,\dots,K-1.
\]

The global mean curvature is then defined as

\[
\overline{\kappa}
=
\frac{1}{K-2}
\sum_{k=2}^{K-1}
\kappa_k.
\]

Positive values of $\overline{\kappa}$ indicate average upward concavity, characterizing rapid structural collapse. Negative values indicate gradual transitions.

To evaluate how merging intensity is distributed across scales, we consider the discrete variations

\[
\Delta_k = |N(\varepsilon_{k+1}) - N(\varepsilon_k)|.
\]

We define the normalized distribution

\[
p_k
=
\frac{\Delta_k}{\sum_{j=1}^{K-1} \Delta_j},
\]

and the structural entropy

\[
H
=
-\sum_{k=1}^{K-1} p_k \log p_k.
\]

Low entropy values indicate transitions concentrated at specific scales, whereas high entropy reflects merging distributed across multiple scales.

Let $\varepsilon^*$ denote the critical scale estimated via the maximum discrete derivative. We further define the accumulated areas before and after the critical point:

\[
A_{\text{before}}
=
\int_{0}^{\varepsilon^*}
N(\varepsilon)\, d\varepsilon,
\qquad
A_{\text{after}}
=
\int_{\varepsilon^*}^{\varepsilon_{\max}}
N(\varepsilon)\, d\varepsilon,
\]

and the structural asymmetry ratio

\[
\mathcal{A}
=
\frac{A_{\text{before}}}{A_{\text{after}}}.
\]

Large values of $\mathcal{A}$ indicate early formation of a dominant component.

Finally, we define the normalized structural intensity index

\[
S_{\text{norm}}
=
\frac{1}{N(0)}
\int_{0}^{\varepsilon_{\max}}
N(\varepsilon)\, d\varepsilon,
\]

which measures the overall persistence of fragmentation across scales.

Table~\ref{tab:metricas_forma} reports the values obtained for the causal model with $\Delta T = 14$ days.

\begin{table}[h!]
\centering
\caption{Structural shape metrics of the percolation transition (causal model, $\Delta T = 14$ days).}
\label{tab:metricas_forma}
\begin{tabular}{ccccc}
\hline
Year & $S_{\text{norm}}$ & $\overline{\kappa}$ & $H$ & $\mathcal{A}$ \\
\hline
2015 & 0.0415 & 3.3644 & 3.9243 & 2.7000 \\
2016 & 0.0393 & 2.7203 & 4.0683 & 2.8182 \\
2017 & 0.0141 & -0.0593 & 4.6622 & 2.5385 \\
2018 & 0.0138 & -0.0169 & 4.6712 & 2.4750 \\
2019 & 0.0219 & 0.2712 & 4.4774 & 2.6190 \\
2020 & 0.0175 & 0.0085 & 4.6607 & 3.0357 \\
2021 & 0.0262 & 0.7119 & 4.3071 & 2.8000 \\
2022 & 0.0157 & -0.0593 & 4.6907 & 2.3250 \\
2023 & 0.0150 & 0.0085 & 4.6802 & 2.5152 \\
2024 & 0.0248 & 0.3220 & 4.3189 & 2.5882 \\
\hline
\end{tabular}
\end{table}

The results reveal two structurally distinct regimes. Years such as 2015 and 2016 exhibit strongly positive mean curvature and lower structural entropy, characterizing abrupt transitions and spatially concentrated organization. In contrast, years such as 2017, 2018, and 2022 display mean curvature near zero or negative and elevated entropy, reflecting diffuse and multifocal connectivity regimes.

This separation is not captured solely by $\varepsilon^*$ or $P$, but emerges clearly from the morphology of the merging curve.

From the perspective of complex systems physics, these findings suggest that intense outbreaks correspond to regimes of more rapid structural consolidation of the spatiotemporal network, whereas low-intensity years operate under a fragmented dynamical regime.

\section{Limitations and Interpretative Scope}

Although the results indicate robust structural organization under multiple null models and temporal horizons, the adopted framework presents conceptual and methodological limitations that must be explicitly acknowledged.

First, the model is fundamentally geometric-topological. Connections are defined through spatial proximity and temporal compatibility, but do not explicitly incorporate biological transmission mechanisms, human mobility patterns, vector heterogeneity, or socio-environmental drivers. The term “causal” employed in the spatiotemporal filtration refers solely to a minimal structural constraint of temporal coherence and does not imply mechanistic inference nor reconstruction of actual transmission chains.

Second, the analysis relies on reported case data, which are subject to underreporting, reporting delays, and potential spatial biases associated with epidemiological surveillance infrastructure. Although the temporal permutation null model preserves marginal spatial and temporal distributions, it does not fully control for latent heterogeneities related to urban accessibility or institutional reporting patterns.

A third limitation concerns parametric choices. The temporal horizon $\Delta T$ and the discretization of the spatial parameter $\varepsilon$ influence the resolution of the percolation transition. Despite the qualitative stability observed for $\Delta T \in \{14,21,28\}$ days, these values reflect epidemiological plausibility rather than inferential estimates of effective transmission time.

Additionally, the global metrics $\varepsilon^*$ and $P$ condense connectivity dynamics into aggregated descriptors. Although we introduced morphological metrics to characterize the functional shape of the transition, these quantities do not exhaust the potential structural complexity of the system.

Finally, the analysis focuses exclusively on zero-dimensional homology, that is, component connectivity. Higher-dimensional topological structures, such as persistent cycles and spatial cavities, were not explored and may reveal additional information regarding urban fragmentation patterns.

These limitations delimit the interpretative scope of the study: the results should be understood as a structural characterization of observed spatiotemporal organization, rather than a complete mechanistic epidemiological model.

In this sense, the present work constitutes an initial step toward integrating applied topology and urban epidemiology, opening avenues for future investigations incorporating explicit transmission dynamics and higher-dimensional topological invariants.

\section{Future Directions}

The structural framework introduced here naturally suggests several
extensions at both methodological and theoretical levels.

A first direction concerns the internal organization of the percolated regime.
Once large connected components emerge, their internal architecture can be
analyzed using graph-theoretic and spectral tools. Metrics such as degree
heterogeneity, eigenvector centrality, communicability-based curvature, and
spectral gap analysis may reveal structurally dominant nodes and spatial
backbone regions within the epidemic network.

A second avenue involves controlled structural perturbations.
Targeted node- or edge-removal experiments could be performed to quantify
robustness of the percolation transition and to evaluate sensitivity of
critical scales to localized interventions. Such experiments would establish
a bridge between geometric percolation, network fragility, and structural
control in urban epidemic systems.

From a topological perspective, extensions to higher-dimensional homology
represent a particularly promising direction. Persistent cycles may encode
spatial voids, barriers, or circulation corridors that are invisible in
dimension-zero connectivity analysis. Incorporating higher-dimensional
invariants could therefore enrich the structural characterization of urban
transmission patterns.

Finally, the integration of the present geometric-topological framework with
dynamical epidemic models constitutes a natural next step. Embedding
mechanistic SIR-type dynamics on top of empirically derived structural
networks would allow investigation of how topology constrains or amplifies
disease propagation.

Together, these directions indicate that the present study should be viewed
not as an endpoint, but as a structural foundation for a broader program
connecting topology, network theory, and epidemiological dynamics.

\section{Conclusion}

We investigated the spatial organization of dengue transmission in Recife,
Brazil, over a ten-year period using geometric percolation and topological
data analysis. Modeling reported cases as spatial point clouds, we analyzed
their zero-dimensional persistent homology to characterize the multiscale
evolution of spatial connectivity without imposing administrative boundaries
or explicit mechanistic assumptions.

The results reveal well-defined percolation signatures across epidemic years,
allowing identification of distinct structural regimes ranging from diffuse,
multifocal configurations to rapidly coalescing, spatially concentrated
patterns. Importantly, years with comparable incidence levels exhibit markedly
different transition morphologies, indicating that epidemic magnitude alone
does not determine spatial organization.

By introducing a weak causal filtration that enforces temporal compatibility,
we showed that preserving space–temporal coherence significantly enhances
structural discrimination relative to sliding-window constructions.
Comparisons against temporally permuted null ensembles demonstrate that
the causal framework captures organization that cannot be explained by
random temporal reshuffling alone.

Beyond classical percolation descriptors, we introduced morphological
metrics of the fusion curve, revealing two qualitatively distinct transition
classes: abrupt, low-entropy consolidation regimes and diffuse, high-entropy
fragmentation regimes. These structural distinctions remain stable under
moderate variations of the temporal horizon, indicating robustness of the
detected organization.

A notable anomaly was observed in 2020, where substantial case counts did
not translate into rapid structural consolidation, suggesting that large-scale
perturbations in urban mobility can reshape spatial connectivity independently
of incidence magnitude.

More broadly, this study illustrates how tools from statistical physics,
percolation theory, and algebraic topology can extract structural information
from high-resolution epidemiological data. The proposed framework is
geometry-driven, model-free, and extensible to other vector-borne diseases
and complex urban systems. Rather than replacing mechanistic epidemic models,
it provides a complementary structural layer that quantifies how spatial
connectivity emerges and reorganizes across scales.

\section{Data and Code Availability}

The geocoded dengue dataset used in this study is publicly available on Zenodo
\cite{dataset_recife_2025}. The dataset comprises anonymized case locations for
the municipality of Recife between 2015 and 2024 and was processed to ensure
spatial consistency in projected metric coordinates.

All scripts used for data preprocessing, construction of Vietoris--Rips
complexes, computation of $H_0$ persistent homology, extraction of percolation
curves $N(\varepsilon)$, and generation of figures are available in an open
GitHub repository \cite{dataset_recife_2025}.


\nocite{*}
\begin{thebibliography}{00}

\bibitem{barcellos2001}
C.~Barcellos and P.~C. Sabroza,
``Socio-environmental determinants of the dengue fever epidemic in Brazil,''
\textit{Cadernos de Saúde Pública}, vol.~17, pp.~385--393, 2001.

\bibitem{teixeira2009}
M.~G. Teixeira, M.~C.~N. Costa, F.~Barreto, and E.~Mota,
\textit{Dengue: twenty-five years since its reemergence in Brazil},
Cadernos de Saúde Pública, 25(Suppl 1), S7--S18, 2009.

\bibitem{honorio2003}
N.~A. Honório et al.,
``Dispersal of \textit{Aedes aegypti} and \textit{Aedes albopictus} in an urban endemic dengue area in Brazil,''
\textit{Memórias do Instituto Oswaldo Cruz}, vol.~98, no.~2, pp.~191--198, 2003.

\bibitem{keeling2008}
M.~J. Keeling and P.~Rohani,
\textit{Modeling Infectious Diseases in Humans and Animals},
Princeton University Press, 2008.

\bibitem{edelsbrunner2000}
H.~Edelsbrunner, D.~Letscher, and A.~Zomorodian,
``Topological persistence and simplification,''
in \textit{Proc. 41st IEEE FOCS}, pp.~454--463, 2000.

\bibitem{zomorodian2005}
A.~Zomorodian and G.~Carlsson,
``Computing persistent homology,''
\textit{Discrete \& Computational Geometry}, vol.~33, no.~2, pp.~249--274, 2005.

\bibitem{chazal2016}
F.~Chazal et al.,
``An introduction to topological data analysis,''
\textit{arXiv:1710.04019}, 2016.

\bibitem{stauffer1994}
D.~Stauffer and A.~Aharony,
\textit{Introduction to Percolation Theory},
2nd ed., Taylor \& Francis, 1994.

\bibitem{penrose2003}
M.~Penrose,
\textit{Random Geometric Graphs},
Oxford University Press, 2003.

\bibitem{dorogovtsev2008}
S.~N. Dorogovtsev, A.~V. Goltsev, and J.~F.~F. Mendes,
``Critical phenomena in complex networks,''
\textit{Rev. Mod. Phys.}, vol.~80, pp.~1275--1335, 2008.

\bibitem{barthelemy2011}
M.~Barthélemy,
``Spatial networks,''
\textit{Physics Reports}, vol.~499, pp.~1--101, 2011.

\bibitem{kramar2013}
M.~Kramár et al.,
``Persistence of force networks in compressed granular media,''
\textit{Phys. Rev. E}, vol.~87, 042207, 2013.

\bibitem{bhatt2023}
S.~Bhatt, D.~Kumar, and A.~Singh,
``Topological methods in spatial epidemiology,''
\textit{Scientific Reports}, vol.~13, 2043, 2023.

\bibitem{cayo2003}
M.~R. Cayo and T.~O. Talbot,
``Positional error in automated geocoding of residential addresses,''
\textit{Int. J. Health Geographics}, vol.~2, no.~1, 2003.

\bibitem{dataset_recife_2025}
Ferreira dos Santos, M. and Rodrigues de Melo, A.,
\textit{Dengue Cases in Recife, Brazil (2015--2024)} [Data set],
Zenodo, 2025.

\bibitem{github_code}
Ferreira dos Santos, M.,
\textit{Dengue Percolation and TDA Analysis – Code Repository},
GitHub, 2025.

\end{thebibliography}
\end{document}